\documentclass[12pt]{article}
\usepackage{graphicx}
\usepackage{latexsym,amsfonts,citesort}
\def\a{\alpha}
\def\b{\beta}
\def\g{\gamma}
\def\d{\delta}

\def\h{\eta}
\def\l{\lambda}

\def\m{\mu}
\def\n{\nu}
\def\r{\rho}
\def\o{\omega}
\def\s{\sigma}

\def\pa{\partial}

\def\be{\begin{equation}}
\def\ee{\end{equation}}
\def\beq{\begin{eqnarray}}
\def\eeq{\end{eqnarray}}
\def\nn{\nonumber}

\newcommand{\bqn}{\begin{eqnarray}}\newcommand{\eqn}{\end{eqnarray}}

\newtheorem{lemma}{Lemma}[subsection]

\begin{document}

\begin{titlepage}
\begin{flushright}
ULB-TH/05-07 \\
\end{flushright}
\vskip 1.0cm

\begin{centering}

{\huge {\bf Lovelock Terms  and BRST \\ \vspace{.2cm} Cohomology}}

\vspace{1cm}

{\large
Sandrine Cnockaert$^{a, }$\footnote{``Aspirant du F.N.R.S., Belgium''} and Marc Henneaux$^{a,b}$}\\
\vspace{.7cm} {\small $^a$ Physique Th\'eorique et Math\'ematique,
Universit\'e Libre de Bruxelles \& \\International Solvay
Institutes, \\ U.L.B.
Campus Plaine C.P. 231, B-1050, Bruxelles, Belgium \\
\vspace{.2cm} $^b$ Centro de Estudios Cient\'{\i}ficos, Casilla
1469, Valdivia, Chile}

\vspace{.5cm}

\end{centering}

\begin{abstract}
Lovelock terms are polynomial scalar densities in the Riemann
curvature tensor that have the remarkable property that their
Euler-Lagrange derivatives contain derivatives of the metric of
order not higher than two (while generic polynomial scalar densities
lead to Euler-Lagrange derivatives with derivatives of the metric of
order four). A characteristic feature of Lovelock terms is that
their first nonvanishing term in the expansion $g_{\l \m} =
\eta_{\l \m} + h_{\l \m}$ of the metric around flat space is a total
derivative. In this paper, we investigate generalized Lovelock terms
defined as polynomial scalar densities in the Riemann curvature
tensor {\em and} its covariant derivatives (of arbitrarily high but
finite order) such that their first nonvanishing term in the
expansion of the metric around flat space is a total derivative.
This is done by reformulating the problem as a BRST cohomological
one and by using cohomological tools. We determine all the
generalized Lovelock terms. We find, in fact, that the class of
nontrivial generalized Lovelock terms contains only the usual ones.
Allowing covariant derivatives of the Riemann tensor does not lead
to new structure. Our work provides a novel algebraic understanding
of the Lovelock terms in the context of BRST cohomology.
\end{abstract}

\vfill
\end{titlepage}

\section{Introduction}
\setcounter{equation}{0} \setcounter{theorem}{0}
\setcounter{lemma}{0} Lovelock terms are polynomial scalar
densities in the Riemann curvature tensor (with indices saturated
with $g^{\m \n}$) \be \sqrt{-g} \, P(R_{\a \b \g \d})
\label{first}\ee that have the remarkable property that their
Euler-Lagrange derivatives contain derivatives of the metric of
order not higher than two. By contrast, generic polynomial scalar
densities lead to Euler-Lagrange derivatives with derivatives of
the metric of order four. The most famous Lovelock term is
probably the Einstein-Hilbert term itself, \be a_{EH} = \sqrt{-g}
\, R\,,\ee whose Euler-Lagrange derivatives yield the Einstein
tensor. Lovelock terms have a long history \cite{Lanczos} and have been systematically determined in all
dimensions in \cite{Lovelock1,Lovelock2}. They have been
considered as possible modifications of the Einstein-Hilbert
Lagrangian in various contexts \cite{Zwiebach,Zumino,TZ} and lead,
in particular, to black hole solutions with interesting properties
\cite{TZ,MS,TrZ,FG}. More recently, the quartic Lovelock term 
has been found to play an important role in deciphering the 
quantum correction structure of M-theory \cite{Green:1981ya,Tseytlin:2000sf,Peeters:2000qj,Deser:2000xz,Damour:2005zb}.

A characteristic feature of any Lovelock term $a$ is that if one
expands it according to the order of the field $h_{\m \n}$ and its
derivatives, \be a = a_k + a_{k+1} + \cdots \; , \ee then the
first nonvanishing term $a_k$ is a total derivative
\cite{Zumino}, \be a_k =
\partial_\m V_k^\m \,.\label{total}\ee Here, each term in the
expansion has definite polynomial order $j$, \be N\, a_j = j \,
a_j \ee where the operator counting the polynomial order is
defined by \be N = h_{\m \n} \frac{\partial}{\partial h_{\m \n}} +
\partial_\r h_{\m \n} \frac{\partial}{\partial (\partial_\r h_{\m
\n})} + \partial_\r \partial_\s h_{\m \n} \frac{\partial}{\partial
(\partial_\r \partial_\s h_{\m \n})} + \cdots \ee The field $h_{\m
\n}$ is the deviation of the metric $g_{\m \n}$ from the Minkowski
metric $\eta_{\m \n}$, \be g_{\m \n} = \eta_{\m \n} + h_{\m \n}
\ee

That the property (\ref{total}) must hold if $a$ is a Lovelock
term is easy to see since $a_k$ is a polynomial in the linearized
curvatures $K_{\a \b \g \d}$ (obtained by setting $g = -1$, $g_{\m
\n} = \eta_{\m \n}$ in the bare metrics involved in (\ref{first})
and by replacing $R_{\a \b \g \d}$ by $K_{\a \b \g \d}$) and thus
reads \be a_k = A^{\m_1 \m_2 \cdots \m_{2k-1} \m_{2k}\n_1 \n_2
\cdots \n_{2k-1} \n_{2k}} \;
\partial^2_{\m_1 \m_2} h_{\n_1 \n_2} \cdots \;
\partial^2_{\m_{2k-1} \m_{2k}} h_{\n_{2k-1} \n_{2k}}\ee for some
constant coefficients $A^{\m_1 \m_2 \cdots \m_{2k-1} \m_{2k}\n_1
\n_2 \cdots \n_{2k-1} \n_{2k}}$. We recall that \be K_{\a \b \g
\d} =-K_{\b \a \g\d} =-K_{\a \b \d\g}= K_{\g\d\a \b }
=\pa^2_{\d [\a}h_{\b] \g}-\pa^2_{\g [\a}h_{\b] \d}
\,,\ee where brackets denote complete antisymmetrization with
weight one. The Euler-Lagrange derivatives of $a_k$ are \be
\frac{\delta a_k}{\delta h_{\r \s}} =
\partial^2_{\m \n}\left(\frac{\partial a_k}{\partial
(\partial^2_{\m \n} h_{\r \s})} \right) \ee and involve terms of
order $k-1$ of the form $\partial^3 h \, \partial^3 h \,
\partial^2 h \, \cdots \, \partial^2 h$, which are quadratic in
the third derivatives of the metric, as well as terms of the form
$\partial^4 h \,
\partial^2 h \, \cdots \,
\partial^2 h$, which are linear in the fourth derivatives. Being of
polynomial order $k-1$, these terms cannot be cancelled by the
contributions coming from the variational derivatives of the
higher order terms $a_{j}$ with $j>k$ since these contributions
are of polynomial order $\geq k$. Hence, the Euler-Lagrange
derivatives of $a_k$ must identically vanish, which implies
(\ref{total})\cite{Poinc1,Poinc2,Poinc3}.

Conversely, that the property (\ref{total}) implies that $a$ is a
Lovelock term is a direct consequence of our analysis below. We
can thus define the original Lovelock terms as the polynomial
densities (\ref{first}) in the curvature that have the central
property (\ref{total}).

In this paper, we investigate generalized Lovelock terms defined
by adopting the property (\ref{total}). More precisely, a
(generalized) Lovelock term of order $k$ is a polynomial scalar
density in the Riemann tensor and its covariant derivatives of
finite (but unrestricted) order, \be a = \sqrt{-g} \, P(R_{\a \b
\g \d}, D_\l R_{\a \b \g \d}, \cdots, D_{l_1}D_{\l_2}\dots
D_{\l_m} R_{\a \b \g \d}) \label{density}\ee such that
\begin{itemize} \item $a$ starts at polynomial order $k$ when
expanded in the fields, $$ a = a_k + a_{k+1} + \cdots , \; \; \;
a_k \not=0 ;$$ \item the first term $a_k$ is a total derivative,
$a_k =
\partial_\m V^\m_k$.
\end{itemize} In
(\ref{density}), indices are contracted with the inverse $g^{\a
\b}$ of the spacetime metric while $D_\l$ denotes the covariant
derivative. The property $a_k = \partial_\m V^\m$ is necessary for
the derivatives of the metric of highest expected order to drop
out from the Euler-Lagrange derivatives -- and, as we shall see,
it turns out to be also sufficient.

It is clear from our definition that if $a$ is a Lovelock term of
order $k$, then $a + b^{(k+1)} + D_\m T^\m$, where $b^{(k+1)}$
starts at polynomial order $k+1$ and $T^\m$ is a vector density,
is also a Lovelock term of order $k$ (even if $b^{(k+1)}$ is not a
Lovelock term of order $k+1$). We shall consider two such Lovelock
terms of order $k$ as being equivalent. Starting from the Lovelock
terms of order $1$, one can successively construct the Lovelock
terms of increasing orders $2$, $3$, etc.

Our main result is that there are, in fact, no new Lovelock terms
besides those already derived in \cite{Lovelock1,Lovelock2}, even
if one allows, as here, derivatives of the Riemann tensor.
Accordingly, nontrivial Lovelock terms of order $k$ can be
assumed to be polynomials of order $k$ in the undifferentiated
Riemann tensor and are proportional to \be \sqrt{-g} \, \d^{\m_1
\m_2 \cdots \m_{2k}}_{\n_1 \n_2 \cdots \n_{2k}} R^{\n_1 \n_2}_{\;
\; \; \; \; \; \; \m_1 \m_2} \cdots R^{\n_{2k-1} \n_{2k}}_{\; \;
\; \; \; \; \; \; \; \; \; \m_{2k-1} \m_{2k}} \label{Lov}\ee with
$$ \d^{\m_1 \m_2 \cdots \m_{2k}}_{\n_1 \n_2 \cdots \n_{2k}} =
\d^{[\m_1}_{[\n_1}\d^{\m_2}_{\n_2}\ldots
\d^{\m_{2k}]}_{\n_{2k}]}\,.$$ Note that while the polynomial order
$N$ is not homogeneous, one may assume that the derivative order
$K$ defined by \be K \, P = \sum_s s \,
\partial_{\r_1 \cdots \r_s} h_{\m \n} \frac{\partial P}{\partial
(\partial_{\r_1 \cdots \r_s} h_{\m \n})}, \ee is homogeneous since
the Riemann tensor is homogeneous of derivative order $2$ and the
Euler-Lagrange operator preserves the derivative order. For a
given even derivative order $s=2k$, there is only one nontrivial
Lovelock term of order $k$, namely (\ref{Lov}), while there is no
nontrivial Lovelock term of odd derivative order.

Our approach relies on the formulation of the problem in terms of
BRST cohomology. We show that Lovelock terms define cohomological
classes of $H(\g_0 \vert d)$ of form degree $n$ and ghost number
zero, which are Lorentz-invariant, while non-Lovelock terms define
cohomological classes of $H(\g_0)$. Here, $n$ is the spacetime
dimension, $\g_0$ is the ``longitudinal differential along the
gauge orbits" of the linearized theory acting on the fields and
the ghosts only (and their derivatives but not on the antifields),
\be \g_0 h_{\m \n} = \partial_\m C _\n + \partial_\n C_\m, \; \;
\; \; \g_0 C_\m = 0 \label{defgamma} \ee (which plays a crucial
role in BRST theory \cite{HT,HTbook}) while $d$ is the spacetime
exterior differential. In (\ref{defgamma}), $C^\m$ are the
diffeomorphism ghosts and their index is lowered with the flat
metric, \be C_\m = \eta_{\m \n} C^\n .\ee Standard techniques of
homological algebra as well as known results on the BRST
cohomology for gravity then enable one to completely determine all
the generalized Lovelock terms.  In fact, once the problem is
reformulated cohomologically, the determination of the Lovelock
terms is quite immediate (section \ref{determining}).

We recall that the equivalence classes $[m]$ of $H(\g_0 |d)$ are
defined by the cocycle condition \be \g_0 m + dq = 0 \ee (for some
$q$), with \be m \sim m' \; \hbox{ iff } \; m-m' = \g_0 p + dr \ee
for some $p$ and $r$. As usual, we switch between form notations
($\g_0 m + dq = 0$) and their duals ($\g_0 m + \partial_\m q^\m =
0$ for a $n$-form, $\g_0 m^\m + \partial_\n q^{\m \n}= 0$ with
$q^{\m \n} = - q^{\n \m}$ for a $(n-1)$-form etc).

Our paper is organized as follows. In the next section, we
reformulate the problem as a cohomological problem in terms of the
BRST differential of the free theory. We then solve this
cohomological problem completely, determining thereby all 
nontrivial Lovelock terms (section \ref{determining}).  To that
effect, we use some cohomological results on $H(\g_0 \vert d)$
established in the appendices.  We finally comment on our results
(sections \ref{a1} and \ref{conclusions}).

\section{Formulating the problem as a cohomological problem}
\setcounter{equation}{0} \setcounter{theorem}{0}
\setcounter{lemma}{0} \label{Scohomo} Let $a$ be a polynomial
density in the curvature and its covariant derivatives, as in
(\ref{density}). One has \be \g a =
\partial_\m (C^\m a). \label{basic}\ee Here $\g$ is the
longitudinal differential along the gauge orbits of the full
Einstein theory, \be \g g_{\m \n} = {\cal L}_C g_{\m \n} , \; \;
\g C^\m = C^\r
\partial_\r C^\m \ee where ${\cal L}_C$ is the Lie derivative
along $C^\m$. If one expands (\ref{basic}) according to the
polynomial degree, one gets as first two equations
\begin{eqnarray} && \g_0 a_k = 0 \,, \label{BRST1}\\ && \g_1 a_k + \g_0 a_{k+1} =
\partial_\m (C^\m a_k) \,,\label{BRST2}\end{eqnarray} where $\g = \g_0 + \g_1 +
\cdots$ is the expansion of $\g$ according to the polynomial
degree. We assume that $a_k \not=0$. Thus $a_k$ is a nonvanishing
polynomial in the linearized curvature $K_{\a \b \g \d}$ and its
ordinary derivatives; it contains therefore at least $2k$
derivative operators. As explained in the introduction, we can
assume that each term has definite derivative order $t$, with $t
\geq
2k$, \beq && N(a_j) = j \, a_j , \\
&& K(a_j) = t \, a_j. \eeq {}From now on, we shall call $\g_0$ the
BRST differential (even though it is only a piece of it, but the
full BRST differential will not be encountered in this paper any
more).

Now, let us assume that $a$ is a Lovelock term of order $k$. Then
\be a_k = \partial_\m V^\m_k\ee where $K(V^\m_k) = (t-1) V^\m_k$.
The equations (\ref{BRST1}) and (\ref{BRST2}) become \beq &&
\partial_\m (\g_0 V^\m_k )= 0 \; , \label{Imp1}\\ &&
\g_0 a_{k+1} = \partial_\m T^\m_{k+1} \label{Imp2}\eeq with \be
T^\m_{k+1} = C^\m \partial_\r V^\r_k - \g_1 V^\m_k .\ee It follows
from (\ref{Imp1}) and the triviality of $d$ that \be \g_0 V^\m_k =
\partial_\n V^{\m \n}_{k\vert 1} \label{Imp3}\ee for some $V^{\m
\n}_{k \vert 1}= - V^{\n \m}_{k \vert 1}$ of ghost number one.
Thus, we see that $V^{\m }_{k}$ and $a_{k+1}$ both fulfill the
cocycle condition of the $\g_0$-cohomology modulo $d$, the former
in form degree $n-1$ and the latter in form degree $n$. Their
respective polynomial, derivative and ghost degrees are \beq &&
N(V^\m_k) = k \, V^\m_k, \; K(V^\m_k) = (t-1) \, V^\m_k, \;
gh(V^\m_k) = 0 \\ && N(a_{k+1}) = (k+1) \, a_{k+1}, \; K(a_{k+1})
= t \, a_{k+1}, \; gh(a_{k+1}) = 0.\eeq {}Furthermore, $V^{\m
}_{k}$ cannot be trivial ($V^\m_k = \partial_\n S^{\m \n}$ for
some $S^{\m \n} = - S^{\n \m}$) since otherwise $a_k$ would be
zero. Thus $V^\m_k$ defines a nontrivial cohomological class of
$H^{n-1,0}_{k,t-1}(\g_0 \vert d)$, where in $H^{i,j}_{k,l}$, the
suffices $i$,$j$ and the indices $k$,$l$ are respectively the form
degree, the ghost number, the polynomial degree and the derivative
degree. {}For expressions involving the ghosts, the polynomial and
derivative degrees are respectively extended as \beq && N =
\sum_{s \geq 0}\left(
\partial_{\r_1 \cdots \r_s} h_{\m \n} \frac{\partial }{\partial
(\partial_{\r_1 \cdots \r_s} h_{\m \n})} + \partial_{\r_1 \cdots
\r_s} C_{\m} \frac{\partial }{\partial (\partial_{\r_1 \cdots
\r_s} C_{\m})} \right),
\\ && K = \sum_{s\geq 1} s \,\left(
\partial_{\r_1 \cdots \r_s} h_{\m \n} \frac{\partial }{\partial
(\partial_{\r_1 \cdots \r_s} h_{\m \n})}+ \partial_{\r_1 \cdots
\r_s} C_{\m} \frac{\partial }{\partial (\partial_{\r_1 \cdots
\r_s} C_{\m})} \right).\; \; \eeq We shall see below that,
similarly, $a_{k+1}$ defines a nontrivial element of
$H^{n,0}_{k+1,t}(\g_0 \vert d)$.

\section{Determining $V^\m_k$}
\setcounter{equation}{0} \setcounter{theorem}{0}
\setcounter{lemma}{0} \label{determining}

Our analysis is based on the resolution of the equation
(\ref{Imp3}) for $V^\m_k$. For that purpose, we use the standard
descent techniques, which rely on the triviality of the cohomology
of $d$ in form degree $<n$ (and $\not=0$). One gets from
(\ref{Imp3}) the chain of equations \beq && \g_0 V^\m_k =
\partial_\n V^{\m \n}_{k\vert 1}, \\ && \g_0
V^{\m \n}_{k\vert 1} = \partial_\r V^{\m \n \r}_{k\vert 2} \\
&& \; \; \; \vdots \nonumber \\ && \g_0 V^{\m_1 \cdots
\m_s}_{k\vert s-1} =
\partial_{\m_{s+1}} V^{\m_1 \cdots \m_s \m_{s+1}}_{k\vert s}
\label{nexttolast}\\ && \g_0 V^{\m_1 \cdots \m_s \m_{s+1}}_{k\vert
s} = 0 \label{last}\eeq where all the $V^{\m_1 \cdots \m_j
}_{k\vert j-1}$ are totally antisymmetric in their upper indices.
The descent stops at some ghost number $s$ (and dual form degree
$n-s-1$) since there is no $p$-form with $p<0$. The antisymmetric
tensors have all polynomial degree $k$ and derivative order $t-1
\geq 2k-1$. If the last term $V^{\m_1 \cdots \m_s
\m_{s+1}}_{k\vert s}$ is trivial in $H^{n-s-1,s}_{k,t-1}$, \be
V^{\m_1 \cdots \m_s \m_{s+1}}_{k\vert s} = \g_0 M^{\m_1 \cdots
\m_s \m_{s+1}} +
\partial_{\m_{s+2}} M^{\m_1 \cdots \m_s \m_{s+1} \m_{s+2}},\ee
then one can remove $V^{\m_1 \cdots \m_{s+1}}_{k\vert s}$ by
redefinitions that affect $V^{\m_1 \cdots \m_s }_{k\vert
s-1}$ as $V^{\m_1 \cdots \m_s }_{k\vert s-1} \rightarrow V'^{\m_1
\cdots \m_s }_{k\vert s-1}=
V^{\m_1 \cdots \m_s }_{k\vert s-1} - \partial_{\m_{s+1}}
M^{\m_1 \cdots \m_s \m_{s+1}}$ and do not affect the preceding
$V$'s, and assume the chain stops one step earlier, at $V'^{\m_1
\cdots \m_s }_{k\vert s-1}$, \be \g_0 V'^{\m_1 \cdots \m_s
}_{k\vert s-1} = 0. \ee

In order to analyse further the descent, we need to consider two
cases: (i) $t>2k$, and (ii) $t=2k$.

\subsection{Case $t>2k$}
The analysis of Appendix B shows that if a $\g_0$-cocycle (i.e., a
solution of (\ref{last})) can be lifted as in (\ref{nexttolast})
then it can be expressed as a polynomial in the linearized
curvature two-form with coefficients that involve the
undifferentiated ghosts $C_{\m}$ and their antisymmetrized
derivatives \be H_{\m \n} =
\partial_\m C_\n - \partial_\n C_\m, \ee up to irrelevant trivial
terms. Derivatives of the linearized curvature appear only in the
trivial terms. Now, $V^{\m_1 \cdots \m_s \m_{s+1}}_{k\vert s}$
involves $s$ ghosts and thus only $k-s$ linearized curvatures.
These $k-s$ curvatures take up $2k -2s$ derivatives, leaving
$t-1-(2k-2s) \geq 2s$ derivatives for the $s$ ghosts. But $s$
ghosts can only take at most $s <2s $ ($s>0$) derivatives in a
nontrivial term, which implies that $V^{\m_1 \cdots \m_s
\m_{s+1}}_{k\vert s}$ is trivial for $s\geq 1$. Thus, we can
assume that the descent stops immediately at $\g_0 V^\m_k = 0$,
i.e., that $V^\m_k$ is invariant for the linearized gauge
transformations.

If $V^\m_k$ is gauge invariant, it is a function of the linearized
curvature and its derivatives\footnote{In this equation and in the
equations below, we use the standard notation $f = f([\phi])$ for
a function $f$ of the field $\phi$ and a finite number of its
derivatives.}, \be V^\m_k = V^\m_k([K_{\a\b \g \d}]) \ee The
indices in $V^\m_k$ are contracted with the Lorentz metric
$\eta^{\m \n}$ so that $V^\m_k$ is a Lorentz vector ($a_k$ is a
Lorentz scalar). Let ${\cal V}^\m$ be the vector density \be {\cal
V}^\m = \sqrt{-g} \, V^\m \ee where $V^\m$ is obtained from
$V^\m_k$ by replacing $\eta^{\m \n}$ by $g^{\m \n}$ and the
linearized curvatures $K_{\a\b \g \d}$ and their derivatives by
the Riemann tensor $R_{\a\b \g \d}$ and their covariant
derivatives. It is clear that $a - D_\m \, {\cal V}^\m$ starts
with a term of order $k+1$ since $(D_\m \, {\cal V}^\m)_k =
\partial_\m V^\m_k = a_k$. Accordingly, there is no nontrivial generalized
Lovelock term of order $k$ if the derivative order $t$ is strictly
greater than $2k$.

\subsection{Case $t=2k$}

When $t=2k$, the analysis proceeds as above, but the descent can
be shortened only to two steps, \beq && \g_0 V^\m_k =
\partial_\n V^{\m \n}_{k\vert 1}, \\ && \g_0
V^{\m \n}_{k\vert 1} = 0 \eeq The antisymmetric tensor $V^{\m
\n}_{k\vert 1}$ contains $k-1$ undifferentiated curvatures (which
take up $2k-2$ derivatives) and one $H_{\m \n}$ (which takes up the
remaining derivative).  As explained in the appendix B, it takes the
form \be \label{trentedeux}V^{\m \n}_{k\vert 1} = B^{[\m\n \s_1 \ldots
\s_{2k-2}]}_{\b_1\b_2\r_1 \ldots \r_{2k-2} } K_{\s_1
\s_2}^{\hspace{15pt}\r_1 \r_2} \ldots K_{\s_{2k-3}
\s_{2k-2}}^{\hspace{41pt}\r_{2k-3} \r_{2k-2}} H^{\b_1\b_2} \ee (see
(\ref{SolForV})).  Lorentz invariance forces the constant tensor
$B^{[\m\n \s_1 \ldots \s_{2k-2}]}_{\b_1\b_2\r_1 \ldots \r_{2k-2} }$
to be proportional to an antisymmetric product of Kronecker $\d$'s,
\be \label{trentetrois} B^{[\m\n \s_1 \ldots \s_{2k-2}]}_{\b_1\b_2\r_1 \ldots \r_{2k-2} }
= \a \, \d^{\m\n \s_1 \ldots \s_{2k-2}}_{\b_1\b_2\r_1 \ldots
\r_{2k-2} }\ee where $\a$ is some constant. [If $n = 2k$, there is
another possibility proportional to $\epsilon^{\m \n \s_1 \cdots
\s_{n-2}}$, but it leads to a solution $a$ which is a total
derivative (not just $a_k$, but the whole $a$ is a total derivative
- in fact the Pontryagin class). The term (\ref{trentedeux}) with $B$
given by (\ref{trentetrois}) yields in fact also a total derivative when $n=2k$.]

Thus, $V^{\m \n}_{k\vert 1}$ reads (up to trivial terms) \be V^{\m
\n}_{k\vert 1} = \a \, \d^{\m\n \s_1 \ldots
\s_{2k-2}}_{\b_1\b_2\r_1 \ldots \r_{2k-2} } K_{\s_1
\s_2}^{\hspace{15pt}\r_1 \r_2} \ldots K_{\s_{2k-3}
\s_{2k-2}}^{\hspace{41pt}\r_{2k-3} \r_{2k-2}} H^{\b_1\b_2}\,.\ee
It follows that \be V^\m_k = 2 \, \a \,\d^{\m\n \s_1 \ldots
\s_{2k-2}}_{\b_1\b_2\r_1 \ldots \r_{2k-2}} K_{\s_1
\s_2}^{\hspace{15pt}\r_1 \r_2} \ldots K_{\s_{2k-3}
\s_{2k-2}}^{\hspace{41pt}\r_{2k-3} \r_{2k-2}} \pa^{\b_1}
h^{\b_2}_{\; \; \; \; \n}\label{solforV}\ee up to strictly
invariant terms. These invariant terms can be removed as above so
that we can indeed assume that $V^\m_k$ is given by
(\ref{solforV}).  Computing the divergence of $V^\m_k$ yields then
\be a_k = - \a \, \d^{\s_1 \ldots \s_{2k}}_{\r_1 \ldots \r_{2k}}
K_{\s_1 \s_2}^{\hspace{15pt}\r_1 \r_2} \ldots K_{\s_{2k-1}
\s_{2k}}^{\hspace{41pt}\r_{2k-1} \r_{2k}}\,,\ee which is the
linearization of the standard Lovelock term of order $k$.
Covariantizing leads to \be a = - \a \, \sqrt{-g} \; \d^{\s_1
\ldots \s_{2k}}_{\r_1 \ldots \r_{2k}} R_{\s_1
\s_2}^{\hspace{15pt}\r_1 \r_2} \ldots R_{\s_{2k-1}
\s_{2k}}^{\hspace{41pt}\r_{2k-1} \r_{2k}}\ee up to terms of order
$>k$.  This term is the complete Lovelock term of order $k$
\cite{Lovelock1,Lovelock2}.  It follows that the standard Lovelock
terms exhaust all the possible Lovelock terms.

\section{Comments on $a_{k+1}$}
\setcounter{equation}{0} \setcounter{theorem}{0}
\setcounter{lemma}{0} \label{a1} We have found all the Lovelock
terms by solving Eq. (\ref{Imp3}), which expresses that $V^\m_k$
is a cocycle of $H(\g_0 \vert d)$.  One can alternatively focus on
Eq.(\ref{Imp2}) and determine $a_{k+1}$, which is also a cocycle
of $H(\g_0 \vert d)$.  The interest of $a_{k+1}$ is that it
coincides with the Pauli-Fierz Lagrangian for the Lovelock term of
order one (in which case $a_1$ is a total derivative and $a_2 =
{\cal L}_{PF}$) and provides generalizations of the Pauli-Fierz
Lagrangian for higher $k$'s.

We shall only sketch here how the analysis proceeds, without
giving all the details. Nontrivial $a_{k+1}$'s exist only if $t =
2k$ as can be seen by examining the descent associated with
$a_{k+1}$, namely, $\g_0 a_{k+1} = \partial_\m T^\m_{k+1}$, $\g_0
T^\m_{k+1} = \partial_\n T^{\m \n}_{k+1}$, etc, and using derivative
counting arguments similar to the ones used above. Furthermore,
one finds that when $t = 2k$, the descent must stop after one
step, \beq && \g_0 a_{k+1} =
\partial_\m T^\m_{k+1}, \\ && \g_0 T^\m_{k+1} = 0. \eeq  The term
$T^\m_{k+1}$ has ghost number one and contains $k$ curvatures;
hence it can only involve the undifferentiated ghost $C_\m$. Using
Lorentz invariance, one gets \be T^\m_{k+1} = \b \, \d^{\m \s_1
\ldots \s_{2k}}_{\n \r_1 \ldots \r_{2k} } K_{\s_1
\s_2}^{\hspace{15pt}\r_1 \r_2} \ldots K_{\s_{2k-1}
\s_{2k}}^{\hspace{41pt}\r_{2k-1} \r_{2k}} C^{\n} \, , \ee which
yields \be a_{k+1} = \b \, \frac{1}{2} \d^{\m \s_1 \ldots \s_{2k}}_{\n \r_1
\ldots \r_{2k} } K_{\s_1 \s_2}^{\hspace{15pt}\r_1 \r_2} \ldots
K_{\s_{2k-1} \s_{2k}}^{\hspace{41pt}\r_{2k-1} \r_{2k}} h_\m^{\;
\n} \, .\ee  One may rewrite $a_{k+1}$ in the more suggestive form
\be a_{k+1} = \b \, {\cal G}^{\m \n}_k \, h_{\m \n} \ee where the
tensor ${\cal G}^{\m \n}_k$ is given by \be {\cal G}^{\m \n}_k =
\frac{1}{2}  \, \d^{\m \s_1 \ldots \s_{2k}}_{\n \r_1\ldots \r_{2k} }
K_{\s_1 \s_2}^{\hspace{15pt}\r_1 \r_2} \ldots
K_{\s_{2k-1} \s_{2k}}^{\hspace{41pt}\r_{2k-1} \r_{2k}}
\ee and fulfills \be {\cal G}^{\m \n}_k = {\cal G}^{\n \m}_k,\; \;
\; \; \; \partial_\m {\cal G}^{\m \n}_k = 0 \ee (identically).

Note also that \be \frac{\delta a_{k+1}}{\delta h_{\a \b}} = \b
(k+1)\, {\cal G}^{\a \b}_k \ee so that \be \frac{\delta a}{\delta
g_{\a \b}} = - \frac{\a}{2} \, \sqrt{-g} \, \Xi^{\a \b } \ee where $\Xi^{\a
\b }$ is the covariantization of ${\cal G}^{\a \b}_k$.  One must
take $\b= -\frac{\a}{2\, (k+1)}$ to match the normalization of $a$
adopted in the preceding section.

\section{Conclusions}
\setcounter{equation}{0} \setcounter{theorem}{0}
\setcounter{lemma}{0} \label{conclusions} In this paper, we have
studied the Lovelock terms of order $k$, defined as polynomial
densities in the curvature tensor and its covariant derivatives
such that the first term in an expansion around flat space is a
total derivative. We have found that Lovelock terms may be assumed
to involve only the undifferentiated Riemann tensor and are thus
exhausted by the standard Lovelock solutions
\cite{Lovelock1,Lovelock2}.  Allowing covariant derivatives of the
curvature tensor does not lead to new solutions.  This is perhaps
not too surprising in view of the topological interpretation of
the Lovelock terms, which can be viewed as the dimensional
continuation of characteristic classes \cite{Zumino}.

Our approach is based on a BRST cohomological reformulation of the
problem and provides a new light on the significance of the Lovelock
terms.  In the full theory, there is not much difference between
Lovelock terms and generic polynomials in the curvature from the
invariance point of view, since all define nontrivial elements of
$H(\g \vert d)$.  However, when restricted to the linear theory,
there is a clear difference between the two: the former define nontrivial 
elements of $H(\g_0 \vert d)$, while the latter define nontrivial 
elements of $H(\g_0)$.  In that respect, it is of interest
to point out that the cohomological groups $H(\g_0 \vert d)$ for
linearized gravity have not been computed yet in all ghost numbers
and ghost degrees.  [These groups are known for $p$-forms
\cite{Poinc2,bv2}, but not for more general tensor fields. They are
important for determining Lagrangians for higher spin gauge fields,
see e.g. \cite{higher}.]

It is of interest to note that a similar property holds for Chern-Simons
terms\footnote{We thank Stanley Deser for stressing this point to us.}: there
is no generalization of the Chern-Simons terms involving higher order derivatives of
the potentials that cannot be reexpressed as a local function of the field strengths
and their derivatives \cite{Poinc2,bv2,BBH85b,DJ99}.  Adding derivatives does not lead
therefore to new structure in the Chern-Simons context either.

\section*{Acknowledgements}
This work is supported in part by the ``Interuniversity Attraction
Poles Programme -- Belgian Science Policy '', by IISN-Belgium
(convention 4.4505.86) and by the European Commission FP6
programme MRTN-CT-2004-005104, in which the authors are associated
to the V.U.Brussel (Belgium).

\appendix
\section{Appendix A: Covariant Poincar\'e lemma for linearized gravity}
\setcounter{equation}{0} \setcounter{theorem}{0}
\setcounter{lemma}{0} \label{appun}

\subsection{A useful preliminary lemma}
Let us first recall a useful lemma proved in \cite{Brandt:1989et}.
\begin{lemma}\label{lemfond}
Let $\eta$ be a differential form depending only on the
derivatives of a field (or set of fields) $\Phi$ and arbitrarily
on a field (or set of fields) $\Psi$. Then \bqn \eta ([\pa \Phi
],[\Psi])=d \o([\Phi ],[\Psi]) \Leftrightarrow \eta = d \Omega
([\pa \Phi ],[\Psi])+ \tilde\eta(d \Phi) \;,\hspace{.5cm}
\tilde\eta (0)=0 \,. \nonumber \eqn
\end{lemma}
In this equation, $\tilde\eta(d \Phi)$ is an exterior polynomial
in the $d \Phi$'s, with coefficients that might involve the $dx$'s
(but we assume no explicit $x$-dependence).
\\

\subsection{Covariant Poincar\'e lemma}
The covariant Poincar\'e lemma for linearized gravity has been
stated first in the fundamental reference \cite{Brandt:1989et},
where the BRST cohomology for full gravity was investigated (without
antifields).

\vspace{.2cm} \noindent {\bf Covariant Poincar\'e Lemma:}
\vspace{.3cm}

\noindent {\it Let $\eta$ be a p-form. Then} \bqn d \eta([K_{\m \n
\r \s}])=0 \Leftrightarrow \eta = d \Omega ([K_{\m \n \r \s}])+
\tilde\eta (K^2_{\m \n})\,, \eqn {\it where $ K^2_{\m \n}=
\frac{1}{2} K_{\m \n \r \s}dx^{\r} dx^{\s}$ is the linearized
curvature two-form.} Again, $\tilde\eta (K^2_{\m \n})$ is here an
exterior polynomial in the $2$-forms $K^2_{\m \n}$.

\vspace{.3cm}

This lemma has been proved in \cite{Brandt:1989et}.  For
completeness, we shall repeat it here, with one slight
modification: namely we do not work in the vielbein formalism.

{}For $p=0$, the lemma is trivial. Indeed $d \eta([K_{\m \n \r
\s}])=0 $ implies that $\h$ is a constant by the usual Poincar\'e
lemma. The lemma is also trivial when $\eta$ does not involve the
field $h$, so we consider that it is at least linear in $h$.

Let us proceed by induction. We assume that the lemma is true for
all $p'<p$ and show that it is still valid for $p$. By the usual
Poincar\'e lemma, $d \eta=0 $ implies that $\eta=d \o_0 ([h])$.
(The subscript denotes the ghost number.) Acting with $\g_0 $ on
$\eta=d \o_0 $ and using the usual Poincar\'e lemma, one obtains
the following descent of equations (where $\o_0 \equiv \hat\o_0 $)
\bqn
\g_0 \hat{\o}_g &=&d\hat\o_{g+1}\hspace{.5cm}0\leq g < G \;,
\nonumber \\
\g _0 \hat\o_G&=&0\,. \nonumber \eqn The form degree of $\hat\o_g$
is given by $p-g-1$. The chain stops at some stage, either because
some $\hat\o_G $ is $\g_0 $-closed, or because the form degree of
$\hat\o_G $ vanishes, i.e., $G=p-1$.

Noting that $\g_0 \o_0$ contains the ghost $C^{\r}$ only with (at
least) one derivative, Lemma \ref{lemfond} together with $\g_0
\o_0=d\hat\o_1$ implies that $(\g_0 \o_0)([h],[\pa C])=d
\o_1([h],[\pa C])+\tilde\o_1(dC)$, where $\tilde\o_1(0)=0$. One
can iterate this step: $\g_0 \o_0=d\hat\o_1$ and $\g_0 \o_0=d
\o_1([h],[\pa C])+\tilde\o_1(dC)$ imply that $\o_1([h],[\pa
C])=\hat\o_1([h],[C])- C^{\s} \frac{ \pa \tilde\o_1(dC)}{\pa
(dC^{\s}) } +d\Omega_1([h],[C])$. Acting on the latter equation
with $\g_0 $, together with $ \g_0 \hat\o_1=d\hat\o_2$, yields
$\g_0 \o_1([h],[\pa C])=d(\hat\o_2([h],[C])-\g_0
\Omega_1([h],[C]))$. As $\g_0$ introduces only differentiated
ghosts $C$, one can apply Lemma \ref{lemfond} and conclude that
$\g_0 \o_1=d \o_2([h],[\pa C])+\tilde\o_2(dC_{\r})$. The iteration
leads to \bqn
\g_0 \o_g&=&d\o_{g+1}+\tilde\o_{g+1}(dC)\hspace{.5cm}0\leq g <G\;, \nonumber \\
\g_0 \o_G&=&\tilde\o_{G+1}(dC)\;,  \label{desc}\eqn where
$\o_g=\o_g([h],[\pa C])$ for $0\leq g \leq G$ and
$\tilde\o_{g}(0)=0$.

Let us split the operator $N$ as follows: $N=N_{[h]}+N_{[C]}$,
where
$$ N_{[h]}=\sum_s \,\partial_{\r_1 \cdots \r_s} h_{\m \n }
\frac{\partial }{\partial
(\partial_{\r_1 \cdots \r_s} h_{\m \n})} \,,~~~~~ N_{[C]} = \sum_s
\,
\partial_{\r_1 \cdots \r_s} C^{\m } \frac{\partial }{\partial
(\partial_{\r_1 \cdots \r_s} C^{\m })}\,. $$ We also define the
operator $\tilde N=K+N\,.$

Equation $\eta=d \o^0 $ splits into eigenfunctions of $N$ and
$\tilde N$. It is sufficient to consider each eigenfunction separately.
The following relations hold for all $g$ (supposing the quantities
involved to be nonvanishing):
$$N(\h)=N(\o_g)=N(\tilde\o_g)\,, \;\tilde N(\h)=\tilde N(\o_g)+1=\tilde N(\tilde\o_{g+1})\,.$$
As $\h=d\o_0$, $\tilde N(\h)$ must be positive. Furthermore, since $\h$
depends on $h$ only through the linearized curvature which
contains two derivatives, $\tilde N(\h)\geq 3N_{[h]}(\h)=3N(\h)$. It is
now straightforward to see that all $\tilde\o$ vanish. Indeed, if
$\tilde\o_g(dC_{\r})\neq 0$, one would have $$\tilde N(\h)\geq 3N(\h)=3
N(\tilde\o_g)=\frac{3}{2} \tilde N(\tilde\o_g)=\frac{3}{2} \tilde N(\h)\,,$$
which is impossible for $\tilde N(\h)>0$.

The last equation of the descent becomes $\g_0 \o_G=0$. This
implies that $$\o_G=\bar{\o}_G([K_{\m \n \r \s}], H_{\m\n})+\g_0
\Lambda_{G-1}\,,$$ where we have taken into account that $\o_G$
does not depend on the undifferentiated field $C_{\r}$. (As a
reminder, $H(\g_0)$ is generated by the linearized curvature and
its derivatives, $[K_{\m \n \r \s}]$, as well as by the
ghosts $ C^\r$ and $H^{\m\n}$, see e.g. \cite{Bouletal}). For
$G=0$, this proves the lemma.

Let us consider $G>0$. The second to last equation of (\ref{desc})
now reads $\g_0 \o_{G-1}= d \bar{ \o}_G + d\g_0 \Lambda_{G-1}$.
One splits the differential $d$ into a part $\bar{d}$ that acts
only on the fields $h$ and a part $\tilde{d}$ that acts only on
the ghosts. Then, $\tilde{d} \bar{\o}_G$ is $\g_0 $-exact and can
be expressed as $\g_0 Y_{G-1}$. So the equation reads $\g_0
(\o_{G-1}-Y_{G-1}+d\Lambda_{G-1})=\bar{d}\bar{\o}_G $. Both sides
have to vanish separately because the r.h.s. is a product of
nontrivial elements of $H(\g_0 )$ and cannot be $\g_0 $-exact.
Without loss of generality, one can write $\bar{\o}_G =\sum_I
P_I([K])M^I_G(H^{\m\n})$, where $P_I$ is a $(p-G-1)$-form in the
curvature and its derivatives, $M^I$ is a polynomial in
$H^{\m\n}$, and $I$ denotes some set of indices.
$\bar{d}\bar{\o}_G =0$ implies $d P_I([K])=0$. The induction
hypothesis for $p'=p-G-1$ can be used to solve this equation and
the solution yields $ P_I([K])=\tilde\h_I(K^2_{\m \n})+
dn_I([K])$, which implies $\bar{\o}_G=\tilde\h_I M^I_G+d( n_I
M^I_G)- n_I d(M^I_G )=\tilde\h_I M^I_G+ d X_G + \g_0 \tilde
Y_{G-1}\,.$ One finally gets $$\o_G=\tilde\h_I(K^2_{\m
\n})M^I_G(H^{\m\n})\,,$$ up to trivial terms that can be removed
by redefinitions of $\o_G$ and $\o_{G-1}$.

We now claim that $G\leq 1$. Indeed, considering that
$N_{[C]}(\o_G)=G$ , $K(\o_G)=(2N_{[h]}+N_{[C]})(\o_G)$ and
$N_1-N=K$, one has
\bqn K(\o_0)+1 =K(\h)\geq 2 N_{[h]}(\h)=2 N(\h)=2 N(\o_G)=
(2N_{[h]}+2N_{[C]})(\o_G)\nonumber \\
=K(\o_G)+G=K(\o_0)+G\,,\nonumber \eqn which implies $G\leq 1$.
Accordingly, in order to complete the proof, we just need to treat
the case $G=1$, as the case $G=0$ has already been solved.

{}From $\o_1=\frac{1}{2}H_{\m\n}\tilde\h^{\m\n}(K^2)$ follows that
$d\o_1=\g_0 (-dx^{\r}\pa_{\m}h_{\n\r}\tilde\h^{\m\n}(K^2))\,.$ The
most general solution $\o_0$ to $\g_0 \o_0=d\o_1$ is
$\o_0=-dx^{\r}\pa_{\m}h_{\n\r}\tilde\h^{\m\n}(K^2)+\Omega([K_{\m
\n \r \s}])$. This in turn gives $$\h=d
\o_0=K^2_{\m\n}\tilde\h^{\m\n}(K^2)+ d\Omega([K_{\m \n \r
\s}])=\h'(K^2)+d\Omega([K_{\m \n \vert \r \s}])\,,$$ where
$\h'(0)=0$. This completes the proof of the lemma.

\section{Appendix B: Some results on $H(\g_0 |d)$}
\setcounter{equation}{0} \setcounter{theorem}{0}
\setcounter{lemma}{0}

In this appendix, we compute the conditions under which an element
$u$ of $H(\g_0)$ at the bottom of a descent of $H(\g_0 |d)$ can be
lifted once, i.e., satisfies the equation $d u=\g_0 v$.

\subsection{The differential $D$}

{}Following \cite{Bouletal}, let us introduce the operator $D$
such that \bqn D \pa_{\n_1} \ldots \pa_{\n_s}h_{\m \r}&=
&d\pa_{\n_1}
\ldots \pa_{\n_s}h_{\m \r}\,,\;\; for\; all\; s\nonumber \\
D C_{\r}&= &\frac{1}{2}dx^{\n} H_{\n \r}\nonumber \\
D \pa_{\n_1} \ldots \pa_{\n_s}C_{\r}&= &0 \,, \;\;\;for\; s>0
\nonumber \eqn The operator $D$ is a differential. It is equal to
the differential $d$ up to $\g_0$-exact terms. It can be
decomposed into a part $D_0$ acting on $h_{\m \n}$ and its
derivatives, and a part $D_1$ acting on the $C_\m$ and its
derivatives. Let $\{\o^I \}$ be a basis of the
(finite-dimensional) space of the polynomials in $C^\r$ and
$H^{\m\n}$.  One has clearly $D_1\o^I=A^I_{~J}\o^J $ for some
matrices $A^I_{~J}$. A grading is associated with the differential
$D$, the $D$-grading, counting the number of $ H^{\n \r}$. The
operator $D_0$ leaves this grading invariant while $D_1$ raises it
by one. It is useful to note that there is a maximal $D$-degree
$m$ for the $\o^I$, due to the fact that the $H$'s anticommute.

\subsection{``Liftable" cocycles}
Let us now compute the $\g_0$-cocycles $u$ that satisfy $du= \g_0
v$ for some $v$. We will assume that the form degree of $u$ is
smaller than $n$, as top-forms trivially satisfy the equation.
Without loss of generality, one can also assume that $u$ has the
form $u=P_I \, \o^I$, where $P_I = P_I([K_{\a \b \g \d}])$ is a
polynomial in the linearized curvature and its derivatives up to
some finite order (with coefficients that can involve the $dx^\m$)
\cite{Bouletal}. Then \bqn
d u&= &(dP_I)\o^I+ (-)^{\vert P_I \vert} P_I d\o^I\nn\\
&=&(dP_I)\o^I+ (-)^{\vert P_I \vert} P_I D_1\o^I +\g_0(\ldots)\nn\\
&=&(dP_J+ (-)^{\vert P_I \vert} P_I A^I_{~J})\o^J +\g_0(\ldots)\nn
\eqn where $\vert P_I \vert$ denotes the grassmanian parity of
$P_I $. If this expression is to be $\g_0$-exact, then $ (dP_J+
(-)^{\vert P_I \vert} P_I A^I_{~J})\, \o^J $ must vanish, or
equivalently, \be dP_J+ (-)^{\vert P_I \vert} P_I A^I_{~J}=0
\,.\label{eqprinc}\ee

We will expand Eq. (\ref{eqprinc}) according to the
$D$-degree. Let us note that it stops at the finite $D$-degree
$m$. We denote by $\o_i^J$ the basis elements of $D$-degree $i$,
and by $P^i_J$ their coefficient in $u$, so that $u=\sum_i
P^i_J\o_i^J$.

We will prove by induction that all $P^i_J$, $0\leq i \leq m$,
have the form \bqn \label{cond1} P^i_J= (-)^{\vert
\Lambda^{i-1}\vert }\Lambda^{i-1}_I A_{i-1\,J}^I + d
\Lambda^{i}_J+\tilde P^i_J\,, \eqn where
$\Lambda^{i}_J=\Lambda^{i}_J([K], dx)$ is at least linear in the
curvature (and $\Lambda^{-1}_J=0$), and $\tilde P^i_J=\tilde P^i_J
(K^2_{\m\n},dx)$ with $i<m$ is an exterior polynomial in the
curvature $2$-form that satisfies \be \label{cond2}\tilde P^i_J
A_{i\, I}^J\o_{i+1}^I=0\,.\ee

In $D$-degree 0, Eq. (\ref{eqprinc}) reads $d P^0_{J}=0$,
which implies, by the covariant Poincar\'e lemma (see Appendix A),
$P^0_{J}=d \Lambda^0_{J}([K], dx)+ \tilde P^0_{J}(K^2_{\m\n},dx)$,
for some $\Lambda^0_{J}$ at least linear in the curvature. So
$P^0_{J}$ satisfies (\ref{cond1}).

We now prove that if $P^i_J$, with $0\leq i \leq m-1$, satisfies
(\ref{cond1}), then the equation in $D$-degree $i+1$ implies
(\ref{cond2}) for $\tilde P^i_J$ and (\ref{cond1}) for
$P^{i+1}_J$. Indeed, in $D$-degree $i+1$, the equation
(\ref{eqprinc}) reads
$$d P_J^{i+1}-(-)^{\vert \Lambda^i \vert}d \Lambda_{I}^i A^{I}_{i\, J}
+(-)^{\vert \tilde P^i \vert}\tilde{P^i}_{I}A^{I}_{i\, J}=0\,.$$
The first two terms contain at least one differentiated curvature
while the last term contains none. So the last term must vanish
separately, which is exactly (\ref{cond2}) for $\tilde P^i_J$. One
is left with $d(P_J^{i+1}-(-)^{\vert \Lambda^i \vert}\Lambda_{I}^i
A^{I}_{i\, J})=0$, which implies (\ref{cond1}) for $P^{i+1}_J$ by
the covariant Poincar\'e lemma.

\vspace{.2cm}

Inserting the expressions (\ref{cond1}) for $P^i_J$ into
$u=P^i_J\o^J_i$ yields (up to $\g_0$-exact terms)
$$u=\tilde{P}_J\o^J +d(\Lambda_J)\o^J +(-)^{\vert \Lambda_{J}
\vert}\Lambda_{J} D_1\o^J =\tilde{P}_J(K^2_{\m\n},dx)\o^J
+d(\Lambda_J([K], dx)\o^J )\,.$$

To summarize:

\vspace{.3cm}

\noindent {\it In form degree $< n$, the $\g_0$-cocycles $u$ that
can be lifted at least once are given by
$$\tilde{P}_J(K^2_{\m\n},dx)\o^J(C^\r,H^{\m\n})$$ where the exterior
polynomials $\tilde{P}_J$ fulfill the constraint $\tilde{P}_J
D_1\o^J=0$ (where there is no sum over $J$), up to
trivial $\g_0$- and $d$-exact terms.}

\subsection{An application}
The constraint is automatically satisfied if $u$ involves only the
differentiated ghosts $H_{\a \b}$ since $D_1 H_{\a \b} = 0$.  In
form degree $n-2$ and ghost number one (a case needed in the
text), a liftable $u$ involving only $H_{\a \b}$ reads thus \be u
= A^{\m_1 \n_1 \m_2 \n_2 \cdots \m_s \n_s}_{\l_1 \l_2 \cdots \l_f
\; \; \; \a \b} \, K^2_{\m_1 \n_1} \, K^2_{\m_2 \n_2} \cdots
K^2_{\m_s \n_s} \, dx^{\l_1} \cdots dx^{\l_f}\, H^{\a \b}\ee with
$n-2 = 2s + f$. The constant coefficients $A^{\m_1 \n_1 \m_2 \n_2
\cdots \m_s \n_s}_{\l_1 \l_2 \cdots \l_f \; \; \; \a \b}$ are
antisymmetric under the exchange of $\m_i$ with $\n_i$, the exchange
of $\l_i$ with $\l_j$ and the exchange of $\a$ with $\b$, and
symmetric under the exchange of the pair $(\m_i,\n_i)$ with the pair
$(\m_j,\n_j)$.  In terms of the dual antisymmetric tensor $V^{\r
\s}$, this expression reads \be V^{\r \s} = B^{\r \s \g_1 \d_1
\g_2 \d_2 \cdots \g_s \d_s}_{\a \b \m_1 \n_1 \m_2 \n_2 \cdots \m_s
\n_s} K^{\m_1 \n_1}_{\; \; \; \; \; \; \g_1 \d_1} \, K^{\m_2
\n_2}_{\; \; \; \; \; \; \g_2 \d_2} \cdots K^{\m_s \n_s}_{\; \; \;
\; \; \; \g_s \d_s} \, H^{\a \b} \label{SolForV}\ee where the
constants $B^{\r \s \g_1 \d_1 \g_2 \d_2 \cdots \g_s \d_s}_{\a \b
\m_1 \n_1 \m_2 \n_2 \cdots \m_s \n_s}$ ($\sim \varepsilon A$) are
completely antisymmetric in their upper indices, $$B^{\r \s \g_1
\d_1 \g_2 \d_2 \cdots \g_s \d_s}_{\a \b \m_1 \n_1 \m_2 \n_2 \cdots
\m_s \n_s} = B^{[\r \s \g_1 \d_1 \g_2 \d_2 \cdots \g_s \d_s]}_{\a
\b \m_1 \n_1 \m_2 \n_2 \cdots \m_s \n_s},$$ and are antisymmetric
under the exchange of $\a$ with $\b$, and symmetric under the exchange
of the pair $(\m_i,\n_i)$ with the pair $(\m_j,\n_j)$.

\end{document}